# LANGMUIR-BLODGETT MONOLAYERS OF CATIONIC DYES IN THE PRESENCE AND ABSENCE OF CLAY MINERAL LAYERS: $N,N'$-DIOCTADECYL THIACYANINE, OCTADECYL RHODAMINE B AND LAPONITE.

**Syed Arshad Hussain[1] and Robert A. Schoonheydt***

*Center for Surface Chemistry and Catalysis, K.U.Leuven, Kasteelpark Arenberg 23, 3001 Leuven, Belgium.*

*[1] **Permanent address**: Department of Physics, Tripura University, Suryamaninagar-799130, Tripura, India. Email: sa_h153@hotmail.com*

- ***Corresponding author:** Robert A. Schoonheydt, Centre for Surface Chemistry and Catalysis, K.U.Leuven, Kasteelpark Arenberg 23, 3001 Leuven, Belgium. Email: Robert.schoonheydt@biw.kuleuven.be*

**Abstract:**

Langmuir-Blodgett (LB) films of $N,N'$-dioctadecyl thiacyanine perchlorate (NK) and octadecyl rhodamine B chloride (RhB18) and their mixtures in the presence and absence of clay mineral layers were investigated by recording surface pressure – area ($\pi - A$) isotherms and by UV-Vis and fluorescence spectroscopies. The $\pi - A$ isotherms of NK, RhB18 and their mixtures are characteristic for liquid expanded state behaviour with repulsive interactions between the two cationic dyes. In the presence of laponite the $\pi - A$ isotherms show liquid expanded and condensed state behaviour. In laponite dispersions and in monolayers, NK has a strong tendency to aggregate with formation of H- and J-aggregates. The absorption and fluorescence maxima of the monomers in the films are at 435 nm and at 480 nm; H-dimer have an absorption maximum around 410 nm and do not fluoresce. J-dimers are present in all the films with absorption maximum at 461 nm and fluorescence at 463 nm. RhB18 is mainly present as monomers in the LB films with an absorption maximum at 576 nm and fluorescence at 595 nm. Fluorescence resonance energy transfer from NK to RhB18 has been observed in clay dispersions and in films with and without laponite. The optimum condition for NK $\rightarrow$ RhB18 fluorescence energy transfer in the films is 90 mol% NK + 10 mol% RhB18.

## Introduction:

Molecules can be organized at inorganic surfaces, such as the surface of elementary clay mineral layers. This is especially the case for cationic surfactants and cationic dyes, ion exchanged on the clay minerals from aqueous solutions. Two types of interactions are important in these hybrid structures: molecule – surface and molecule – molecule. (i) Adsorption from dilute aqueous solution leads to a strong concentration of the molecules at the surfaces of the elementary clay mineral layers. This local concentration exceeds the equivalent solution concentration for formation of dimers and aggregates. Thus, H- and J-aggregates are easily observed, especially in the case of cationic dyes even at loading of < 1% of the cation exchange capacity (CEC). (ii) Surface-molecule interactions compete with the above mentioned molecule-molecule interactions. They induce spreading of the (cationic) molecules over the surface and therefore counteract aggregation. If surface – molecule interactions predominate, monomers are preferentially observed. This is the case at low water content in clay mineral – cationic dye systems. Molecule-molecule interactions are predominant at high water content and molecular aggregates are easily formed. The extent to which these general rules are valid depends on the type of molecule and the type of clay mineral [1-5].

If the elementary clay mineral layers are organized in one plane, one can in principle realize a two-level organization, that of the molecules at the surface and that of the elementary clay mineral layers themselves. Spin coating [6-8], self assembling (SA) or Layer-by-Layer asembly (LbL) [9-11] and the Langmuir-Blodgett (LB) technique [12-13] have been used to realize this two-level organization. A review has recently been published discussing SA and LB layers of negatively charged elementary clay layers, alternating with (mono) layers of cationic surfactants such as polyelectrolytes, dyes and proteins [14]. The LB technique is preferred because it leads to smoother films with a good organization of elementary clay mineral layers and cationic surfactants, including cationic dyes. These films can have interesting properties, such as second harmonic generation, photo-induced magnetization, electron transfer and chiral sensing [14]. These properties are determined by the specific organization of the adsorbed molecules and of the elementary clay mineral layers in the LB films.

In this communication we report the spectroscopy of mixed molecular assemblies of a thiacyanine dye, 3-octadecyl-2-[(3-octadecyl-2(3H)-benzolylidene) methyl] benzothiazolium perchlorate or $N,N'$-dioctadecyl thiacyanine perchlorate (NK) and a xanthane dye, octadecyl ester of rhodamine B chloride (RhB18), organized in LB monolayers with and without elementary clay mineral layers. Fluorescent cyanine and thiacyanine dyes are widely used as markers and sensor systems [15]. Their absorption and fluorescence spectra can change remarkably upon adsorption [16]. RhB18 has a high absorption coefficient and high quantum yield. It can be used as photosensitizer, as quantum counter and as an active medium in dye lasers [17]. The aim of this study is to investigate the spectroscopic properties of the mixed dye system (NK + RhB18) in a LB monolayer with and without elementary clay mineral layers.

Aggregates of two different dyes in LB films have recently gained importance, because of (i) their specific spectroscopic properties, (ii) the control on the aggregation process in the course of the LB film preparation process. These films might also act as model systems for molecular clusters and crystals [15-16]. The J aggregates of thiacyanine dyes are characterized by a sharp absorption band, red-shifted from the monomer band in the visible region of the spectrum, and by typical fluorescence spectra. Because of remarkable non-linear optical properties, these J aggregates are promising candidates for future technological applications in optoelectronic devices [17]. Mixed aggregates of two different dyes in organized molecular assemblies seem to be useful for construction of an artificial photosynthetic system and for imaging systems [18]. In some cases fluorescence resonance energy transfer (FRET) between the two dyes might occur [18].

Energy transfer between two dyes is an important physical phenomenon with considerable interest in the understanding of some biological systems and with potential applications in optoelectronic and thin film device development [19,20]. The rate and efficiency of the energy transfer process are very sensitive to the fluorescence properties of the individual molecules in the molecular assemblies and to their geometrical arrangement [21]. Thus, differences in arrangement of the donor and acceptor molecules in the LB films without and with clay mineral layers can be observed and discussed in terms of molecular organization.

**Experimental:**

**Materials:**

3- octadecyl-2[3-octadecyl-2 (3H)-benzothizolidene) methyl] benzothiazolium perchlorate or $N,N'$- dioctadecylthiacyanine perchlorate (NK) [Hayashibara Biochemical Laboratories Inc] and octadecyl rhodamine B chloride (RhB18) [Molecular Probes] were used as received. Molecular structures of the dyes are shown in figure 1. The dyes were dissolved in either HPLC grade chloroform (99.9 % Aldrich, stabilized by 0.5-1% ethanol) or HPLC grade methanol [Acros Organics, USA]. The clay mineral used in the present work was Laponite, obtained from Laponite Inorganics, UK and used as received. The size of the clay platelet was less than 0.05 $\mu m$ and its CEC was 0.74 meq/g, determined with CsCl [22].

**Film preparation:**

A commercially available Langmuir-Blodgett (LB) film deposition instrument (NIMA Technology, model 611) was used for isotherm measurement and monolayer film preparation. Either pure Milli-Q water or clay dispersions stirred for 24 h in Milli-Q water were used as subphase. The clay concentration was fixed at 2 mg/L. Solutions of NK, RhB18 as well as NK-RhB18 mixtures at different mole fractions were prepared in chloroform and were spread on the subphase with a microsyringe. For each isotherm measurement 40 $\mu l$ of dye solution in chloroform with a total concentration of $0.5\times10^{-4}$M was spread on the subphase. After 15 minutes (water subphase) and 30 minutes (clay dispersion subphase), the floating monolayer was compressed at a rate of 10 $cm^2min^{-1}$ to monitor the surface pressure-area isotherms. The surface pressure was recorded using a Wilhelmy arrangement. Each isotherm was repeated once and consistent results were obtained. Monolayer films were deposited in upstroke (lifting speed 5 mm $min^{-1}$) at a fixed surface pressure of 15 mN/m onto fluorescence grade quartz plates for spectroscopy and on Si wafers for AFM. The transfer ratio was found to be 0.98 $\pm$ 0.02.

**Characterizations:**

The Atomic force Microscopy (AFM) images of RhB18-laponite hybrid monolayer films was taken in air with a commercial AFM system Autoprobe M5 (Veeco Instr.) using silicon cantilevers with a sharp, high apex ratio tip (UltraLevers$^{TM}$, Veeco Instr.). The AFM image presented here was obtained in intermittent-contact ("tapping") mode. Typical scan areas were 3×3 $\mu m^2$. The monolayers on Si wafer substrates were used for the AFM measurements.

UV-Vis absorption and fluorescence spectra of the solutions and films were recorded with a Perkin Elmer Lambda-12 Spectrophotometer and a SPEX Fluorolog 3-22 double grating Fluorescence Spectrophotometer respectively. The excitation wavelength ($\lambda_{ex}$) was 430 nm. For absorption measurement the films were kept perpendicular to the incident light and a clean quartz slide was used as reference. The fluorescence light was collected from the sample surface at an angle of $45^0$ (front face geometry). In some cases excitation spectra were also recorded with the fluorescence wavelength fixed at 590 nm on a Fluorescence spectrophotometer (LS-55, Perkin Elmer).

**Results:**

**Monolayer Characteristics at the air-water interface**

Figure 2a shows the surface pressure-molecular area ($\pi - A$) isotherms of xRhB18:(1-x)NK in the absence of laponite with molar fraction x = 0, 0.2, 0.5, 0.6, 0.9 and 1.0. The RhB18 isotherm shows a smooth rise of the surface pressure ($\pi$) upon compression of the monolayer. At $\pi$=32.2 $mNm^{-1}$ a phase transition occurs, probably from the liquid expanded state to a mixture of liquid and condensed states. The lift-off area, determined with the method described by Ras et al. [18] is 1.76 $nm^2$. This lift-off area and the molecular areas of 0.85 $nm^2$ and 0.64 $nm^2$ at surface pressures of 20 and 30 $mNm^{-1}$ are in agreement with literature values [18-19]. The NK isotherm is steeper and does not show such a sharp phase transition as the RhB18 isotherm. The lift-off area of NK is 1.13 $nm^2$. Also higher surface pressures can be reached, indicative for the higher stability of the NK monolayer.

The isotherms corresponding to the RhB18-NK mixtures are situated between those of the pure components with lift-off areas ranging in between those of the individual components. Up to x in the range 0.6 - 1 the isotherms are RhB18 like with a relatively sharp phase transition; for x < 0.6 they are NK like.

Figure 2b shows the surface pressure-molecular area ($\pi - A$) isotherms of xRhB18:(1-$x$)NK in the presence of laponite with $x$= 0, 0.1, 0.5, 0.9 and 1.0. These isotherms are different from those in the absence of laponite particles in several aspects. Here the isotherms possess typical shapes of distinct liquid – expanded and condensed solid phases. The latter was not observed for the mixed isotherms in the absence of clay. The transition from liquid expanded to condensed phase occurs at around $\pi$= 15 $mNm^{-1}$ for NK and NK + RhB18 mixtures up to the RhB18 mole fraction of 0.5. For higher RhB18 mole fractions $x$, the surface pressure at which the phase transition occurs increases to 20 $mNm^{-1}$ for $x$ = 0.9 and is 25 $mNm^{-1}$ for the RhB18 – laponite isotherm. The lift-off areas are almost independent of the type of dye and equal to 1.4 – 1.5 $nm^2$. This value is much lower than that of RhB18 and higher than that of NK, both taken in the absence of laponite. They are largely determined by the laponite particles in the monolayer films.

The compressibility (C) of the monolayer at the air-water interface was calculated, according to [20]

$$C = -\frac{1}{a_1}\frac{a_2 - a_1}{\pi_2 - \pi_1}$$

Where $a_1$ and $a_2$ are the areas per molecule at the surface pressures $\pi_1$ and $\pi_2$ respectively. For monolayers with laponite two compressibilities were calculated, one for the liquid expanded part of the ($\pi - A$) isotherm (below 15 $mNm^{-1}$) and one for the condensed phase in the high surface pressure regime (above 30 $mNm^{-1}$). The compressibilities are given in table 1.

In the absence of laponite the compressibility increases with increasing mole fraction of RhB18 with a maximum of 30 $mN^{-1}$ at $x$= 0.5 and a shallow minimum of 26.6 $mN^{-1}$ at $x$= 0.8. For the liquid expanded phase the compressibilities of the laponite-dye composite films at the air-water interface are higher than those of the dyes in the absence of laponite for all dye compositions, except at the highest RhB18 mole fractions (x = 0.9 and 1). The compressibility of the solid condensed phase is small and in the range of 5.6 – 8.6 $mN^{-1}$, except for the RhB18 – laponite system ($x$= 1) with a compressibility of 15.4 $mN^{-1}$.

The miscibility or the phase separation of RhB18 and NK in the LB films can be determined on the basis of the shape of the $\pi - A$ isotherms in the absence of laponite for various mole fractions using the additivity and surface phase rules [20]. In the case of an ideally mixed monolayer RhB18 and NK are randomly distributed and the intermolecular attraction forces are equal: $F_{11} = F_{12} = F_{22}$. In a completely immiscible monolayer, the intermolecular attraction forces among similar molecules ($F_{11}, F_{22}$) are larger than among different molecules ($F_{12}$): $F_{11} > F_{12} < F_{22}$ (1 refers to RhB18; 2 to NK).

The miscibility of a mixed monolayer can be examined by quantitative analysis of the excess area ($A^E$) of the mixed monolayer at the air-water interface. The excess area is obtained by comparing the experimentally observed average area per molecule ($A_{12}$) of the mixed monolayer consisting of the components 1 and 2, with that of an ideally mixed monolayer ($A_{id}$). It is given by $A^E = A_{12} - A_{id}$, with $A_{id} = A_1 x_1 + A_2 x_2$ $A_1$ and $A_2$ are the areas occupied by the monomers of RhB18 and NK respectively and $x_1$ & $x_2$ are the mole fractions of the components in the mixture ($x_1 + x_2 = 1$).

In the ideal case the plot of $A_{12}$ versus $x_1$ will be a straight line. Any deviation from the straight line ($A^E = A_{12} - A_{id} \neq 0$) indicates partial miscibility and non-ideality [20-21]. If attractive intermolecular forces are dominant, $A^E$ will be negative. On the other hand, positive values ($A^E > 0$) indicate repulsive interaction between the constituent components of the mixed monolayer.

The inset of figure 2a presents the plot of the $A^E / A_{id}$ versus the mole fraction of RhB18 ($x_1$) in the RhB18-NK mixed monolayers at surface pressures of 5, 10, 15, 20, 25 and 30 mN/m.

From the figure a noticeable deviation from the additivity rule is observed, which is an indication of interaction among the constituent molecules of the binary mixture in the monolayer. This deviation is predominantly positive, meaning that repulsive interactions are dominant. At RhB18 mole fractions of 0.4 and 0.8 the $A^E / A_{id}$ values are almost independent of the surface pressure. At mole fractions of 0.1, 0.5 and 0.6 there is a large variation of $A^E / A_{id}$ values with surface pressures and $A^E$ is even negative at the lowest surface pressures.

***Atomic Force Microscopy:***

Figure 3 shows a typical AFM image of RhB18-laponite hybrid monolayer deposited on a Si substrate at a surface pressure of 15mN/m and a waiting time of 15 min along with the line analysis spectrum.

In the figure the laponite particles are easily observed. The hybrid monolayer consists of a closely packed array of hybridized laponite particles. The surface coverage exceeds 80%. White spots are indicative of aggregates of laponite particles, black spots indicate uncovered substrate regions. Individual laponite layers are not clearly resolved due to the high layer density. From the height profile analysis it is seen that the height of the monolayer varies between – 2 nm and + 2 nm. This includes the height of individual laponite layer, aggregates of individual layer and RhB18 molecules.

**UV-Vis absorption Spectroscopy**

**Solution and clay dispersions**

The UV-Vis absorption spectra of chloroform solutions of RhB18, NK and their mixtures are characteristics for the presence of monomers (figure not shown). The RhB18 solution absorbs light at 556 nm, which is attributed to the monomeric form of the dye. A small shoulder around 518 nm can be assigned to the $0 \rightarrow 1$ vibronic component. The band maximum of monomeric NK is at 430 nm with the $0 \rightarrow 1$ vibronic component at 406 nm. The absorption spectra of the RhB18-NK mixtures in chloroform are characterized by the same absorption bands. Also, the intensity of these bands is proportional with the molar ratio. This indicates the absence of significant chemical or physical interactions between RhB18 and NK molecules in $CHCl_3$ solution.

The absorption spectra of the dyes in clay dispersions with total dye loading of 10% of CEC of Laponite are shown in figures 4. Marked changes in intensity as well as band profile with respect to the chloroform solution spectra are observed. The NK spectra consist of two bands at 406 nm and 427 nm. The latter is close to the 430 nm band in solution and assigned to the monomer. In the Laponite dispersion the 406 nm band is more intense than in chloroform solution. We ascribe this to an increased contribution of H-dimers.

The RhB18 spectra in the clay dispersions show two bands at 525 nm and at 565 nm. The first is the $0 \rightarrow 1$ vibronic transition of the monomer with – possibly - some contribution from the H-dimer; the second is the $0 \rightarrow 0$ vibronic transition of the monomer of RhB18 at the clay surface. One notices the red shift of 9 nm with respect to the $0 \rightarrow 0$ band in chloroform solution, the clay environment being more polar than the chloroform solvent.

**Langmuir-Blodgett (LB) Films**

Absorption spectra of LB monolayers of the two dyes and their mixtures in absence and presence of laponite are shown in figures 5a and 5b respectively. The spectra of NK are characterised by three bands at 410 nm, 435 nm and 453 nm. The same bands with roughly the same intensity ratios are present when the NK monolayer is deposited at 30 $mNm^{-1}$ (figure not shown). The RhB18 spectra are characterized by the main band at 576 nm and the 525 nm shoulder. At a mole fraction $x$=0.1 of RhB18 the 576 nm band is broadened towards lower wavelength and covers the range 555 – 575 nm.

In the presence of laponite the NK absorption spectra (figure 5b) of the monolayers consist of two bands with maxima around 410 and 435 nm. They have almost equal intensities. The 453 nm component observed in LB films without laponite is not resolved, but might be hidden in the tail of the 435 nm band. However when the (NK + laponite) monolayer is deposited at 30 $mNm^{-1}$ a pronounced, sharp 461 nm absorption is observed (inset of fig. 5b). This is typical for J- aggregates.

In the presence of laponite the RhB18 main absorption band maximum shifts from 565 to 576 nm as the mole fraction increases. No effect of surface pressure on the spectra was observed.

**Fluorescence Spectroscopy**

**Solution and clay dispersions**

The fluorescence spectra of solutions of NK, RhB18 and their mixtures in chloroform (figure not shown) have a broad band with maximum around 483 nm, due to NK fluorescence, and a band at 577 nm of RhB18. The intensity ratios of these bands in the spectra of mixed dyes follow the molar ratio of the two dyes in solutions. The intensity of the NK fluorescence is 6 X smaller than that of RhB18, due to a difference in quantum yield.

The fluorescence spectra of the dyes in laponite dispersions (fig. 6) at a total dye loading of 10% of the CEC the NK fluorescence is characterized by a broad band 450-480 nm.. The RhB18 fluorescence is at 582 nm. Its maximum intensity is observed at RhB18 loading of 10% of total dye and this intensity again exceeds that of NK.

**Langmuir-Blodgett (LB) Films**

The fluorescence spectra of LB films of mixed (NK + RhB18) dyes in the absence and presence of laponite particles are shown in figure 7a and 7b respectively. In the absence of laponite the NK emission consists of two bands, a relatively sharp band at 463 nm and a shoulder around 480 nm. The intensity increases gradually with loading. The RhB18 fluorescence is at 599 nm and maximum intensity is obtained for $x$=0.1 The surface pressure for film deposition has no effect on the spectra of RhB18 (figure not shown).

In the presence of laponite particles the NK fluorescence is characterized by a broad band covering the 460 -480 nm region. At 100% NK the 463 nm fluorescence band becomes well resolved. When the (NK + laponite) monolayer is deposited at 30 $mNm^{-1}$ a sharp NK fluorescence is observed at 463 nm with a shoulder around 480 nm. The 463 nm fluorescence corresponds with the sharp absorption band at 461 nm. Thus the 463 nm band is due to J-dimer fluorescence. The RhB18 fluorescence is at 595 nm with a pronounced maximum intensity, when the RhB18 content is 10% of the total dye content.

Figure 7c shows the fluorescence spectra of LB monolayers containing NK, RhB18 and laponite. These monolayers were prepared with constant amount of NK (20 $\mu l$, $5\times10^{-5}$ M) and varying amounts of RhB18 (5, 10, 20 and 30 $\mu l$, $5\times10^{-5}$ M). In all cases the NK fluorescence is weak and broad, covering the 463 and 480 nm fluorescence of J-aggregates and monomers. The RhB18 fluorescence at 590 nm decreases with increasing amount of RhB18. Also as the RhB18 fluorescence intensity increases, the NK fluorescence decreases. Excitation spectra were recorded with the emission wavelength fixed at 590 nm (see insert of figure 7c). They show two components at 461 and 435 nm, due respectively to J-dimers and monomers of NK.

## Discussion

### Surface pressure – area isotherms

The lift-off areas of NK and RhB18 molecules are respectively 1.13 nm$^2$ and 1.76 nm$^2$. The compressibility of the NK monolayer is less than that of RhB18 monolayer. These observations mean that the NK molecules can be more tightly packed into a rigid monolayer, while the RhB18 monolayer consists of loosely packed molecules which can be more easily compressed than the NK monolayer. In the monolayer of both molecules the lift-off areas exceed those expected for the limiting cases of complete miscibility or complete immiscibility. This observation is indicative for repulsive interactions NK-RhB18, which is not surprising in view of the positive charge of the molecules.

In the presence of laponite the molecular areas of NK and RhB18 are 1.54 nm$^2$ and 1.47 nm$^2$ respectively. The molecular areas are almost independent of the type of the molecule. We conclude that laponite particles are fixed at the air-water interface to form a so – called hybrid monolayer consisting of laponite layers and dye molecules. With a CEC of 0.74 meq/g and an estimated surface area of 750 -850 m$^2$/g, the average area per negative charge is 1.41 nm$^2$ – 1.68 nm$^2$, very close to the lift off areas of 1.54 nm$^2$ and 1.47 nm$^2$ observed for the NK-Laponite and RhB18-Laponite systems respectively. The conclusion is that every cationic dye molecule in the monolayer neutralizes one negative charge on the laponite particles by an ion exchange reaction. The amount of laponite in the films is then estimated to be 2.6 μg under the hypothesis that the film is composed of single layers in a strictly two-dimensional array. The AFM images do not confirm this. They show a film containing a mixture of single laponite layers and aggregates of various sizes. Some of these have a thickness of 4 nm. The thickness of a single layer is 0.96 nm. A hydrated $Na^+$-Laponite layer has a thickness of 1.25 nm. A single Laponite layer with a monomoleucular layer of RhB18 can have a thickness of 1.85 nm [23]. Thus a particle with a thickness of 4 nm might consist of 2, 3 or 4 layers of Laponite, depending on the conditions. The 2.6 μg must then be considered as a lower limit.

### Spectroscopy:

### NK:

The spectroscopic signatures of NK are summarised in table 2. The interpretation of the absorption spectra is straightforward. The monomer band is found at 430 nm, its position being only weakly dependent on the environment. H- dimers absorb around 410 nm. They have been observed in the aqueous laponite dispersions and in the films i.e. in systems in which the NK molecules are concentrated in a small volume. The absorption band of J-dimers or J- aggregates has its maximum at 461 nm. Its is only clearly seen in films with and without clay. In the latter case a film deposited at a surface pressure of 30 $mNm^{-1}$ is necessary. This means that J-aggregate formation is suppressed by adsorption of NK molecules at the laponite surfaces. Their formation at high surface pressure could mean that the NK molecules are desorbed and form regions of J aggregates in the spaces between the laponite layers in the film.

The fluorescence spectra of NK can be interpreted with the fluorescence of the monomer around 480 nm and of the J aggregates at 463 nm. The H dimers or aggregates do not fluoresce. All the fluorescence spectra of NK, except the solution spectra show the presence of monomers and J aggregates in various ratios. J aggregate fluorescence is dominant in aqueous laponite dispersions and in the films without and with laponite. In the latter case this is only the case with 100% NK and no RhB18, confirming our observation on the absorption spectra that J aggregate formation is suppressed by adsorption of NK on the laponite surface. In all cases, except the solution, the monomer fluorescence of NK is weaker than or of the same intensity as the fluorescence of J aggregates.

**RhB octadecyl ester and comparison with NK:**

The spectroscopic signatures of RhB18 are summarized in Table 3. They can be interpreted in terms of monomer absorption and fluorescence. With the spectra of the choroform solutions as references, there are significant shifts to higher wavenumbers in clay dispersions and in films, nicely reflecting the more polar environments. These shifts are more pronounced than in the case of NK. The apparent Stokes' shifts of the monomers are also significantly different. They are in the range 520 – 667 $cm^{-1}$for RhB18 of 2370 – 2550 $cm^{-1}$ for NK.

The differences between both molecules can be quantitatively understood on the basis of their molecular structure (figure 1). For NK resonance structures can be drawn with a positive charge on either of the two N atoms. This means that the two C-C bands connecting the ring systems to the central CH group have double band character. Rotation around these bands is restricted.

The chromophore unit of RhB18 consists of 3 fused six-rings and two resonance structures can also be drawn too [24]. Excitation of a $\pi$ electron of the RhB18 chromophore will not change the conformation of the chromophore, the latter being a rigid, three fused six-rings system. Thus, the Stokes' shift is small, as observed. The NK chromophore is not so rigid, as it consists of two fused five- and six-rings, connected by a $=CH=$ bridge. A larger conformational change is expected upon excitation and is indeed observed, as evidenced by the Stokes' shifts of table 4.

Finally, NK-NK interactions predominate over NK-laponite interactions and in general NK-environment interactions. The reverse is true for RhB18. It follows that NK has a strong aggregation tendency with formation of H- and J-aggregates and its spectroscopic signatures are weakly dependent on the environment. RhB18 has a weak aggregation tendency and its spectroscopic signatures are significantly dependent on the environment.

**NK-RhB18 systems:**

Mixing NK and RhB18 in solution, in aqueous laponite dispersions and in films with and without laponite has no significant effect on the overall intensities of the absoprtion spectra of the two dyes. The intensities increase regularly with increase of the amount of the dye in the systems. Mixing has a significant effect on the intensities of the fluorescence spectra. In the aqueous clay dispersions and in the LB monolayers the fluorescence intensities of NK always increase with increasing molar fraction of NK. The RhB18 fluorescence attains a maximum at a characteristic molar fraction of 0.1. When the amount of RhB18 varies at constant NK loading, maximum intensity of RhB18 fluorescence is observed at the lowest

RhB18 loading. With the emission wavelength fixed at the maximum of the RhB18 fluorescence (590 nm) an excitation spectrum is obtained with the characteristic bands of the monomer and J aggregate of NK. Thus fluorescence resonance energy transfer occurs from NK monomers and J aggregates towards RhB18. This occurs in clay dispersions and in films i.e. whenever the molecules are brought in close proximity.

The complexity of the system is prohibitive for any attempt to quantify the FRET phenomenon and for an in-depth interpretation. Indeed the system contains monomers, H- and J- dimers of NK. Energy transfer between these species might occur. At loadings of RhB18 above a molar fraction of 0.1 self-quenching might occur. Finally there is the question of the distribution of the dyes. Ideal mixing does not occur and the system might contain regions of NK molecules and regions of RhB18 molecules. Thus, future research is aimed at the optimization of the systems in such a way that the two dyes are present in monomeric form only. Anyhow it is reconforting to see that laponite particles are helpful in organizing molecules and in maximizing processes such as fluorescence resonance energy transfer.

**Conclusions:**

Langmuir and Langmuir-Blodgett (LB) films of RhB18, NK and their mixtures were successfully prepared with and without laponite clay mineral layers. $\pi - A$ isotherms of mixed dyes in the absence of laponite particles reveals repulsive interaction between the dyes, $\pi - A$ isotherms of mixed dyes in the presence of laponite sheets confirm the incorporation of laponite layers in the monolayers. The isotherms reveal a liquid expanded state at low surface pressure and a condensed state at high surface pressure.

NK has a strong tendency to aggregate in dilute laponite dispersions and in films with formation of both H and J aggregates. The latter are especially prominent in laponite films deposited at high surface pressure. They are indicative of desorption of NK from the laponite layers with formation of J aggregates in the void regions between the laponite layers. In the system investigated RhB18 is present as monomers. Fluorescence resonance energy transfer from NK monomers and J aggregates to RhB18 monomers has been observed in clay dispersions and in films. A molar ratio RhB18:NK of 0.1 seems to be optimal. However the system is not yet optimized. The adsorption of NK on laponite layers in the films prohibit extensive J aggregate formation. In this way incorporation of clay layers in the films helps in the organization of cationic dye molecules and in optimization of the systems towards a specific property, in the present case fluorescence resonance energy transfer.

**Acknowledgment:**

SAH acknowledges the Fund for Scientific Research – Flanders for a postdoctoral fellowship. This research is funded by IAP- (Inter-University Attraction Pole) and by CECAT (Centre of Excellence – K. U. Leuven).

**Figure caption:**
Figure 1 Molecular structure of (a) RhB18 and (b) NK

Figure 2a: Surface pressure versus molecular area ($\pi - A$) isotherms of xRhB18:(1-x)NK in absence of clay with molar ratio, x = 0, 0.2, 0.5, 0.6, 0.9 and 1.0. The inset shows the plot of $A^E / A_{id}$ as a function of mole fraction of RhB18 in NK.

Figure 2b: Surface pressure versus molecular area ($\pi - A$) isotherms of xRhB18:(1-x)NK in presence of laponite with molar ratio, x = 0, 0.1, 0.5, 0.9 and 1.0.

Figure 3: AFM image of RhB18-laponite hybrid monolayer.

Figure-4 UV-Vis absorption spectra of mixed dyes (RhB18 & NK) in laponite dispersion. Dye loading is 10% of CEC of laponite.

Figure-5a UV-Vis absorption spectra of monolayer LB films of mixed dyes (RhB18 & NK) in absence of laponite deposited at 15 mN/m surface pressure.

Figure 5b UV-Vis absorption spectra of monolayer LB films of mixed dyes (RhB18 & NK) in presence of laponite deposited at 15 mN/m surface pressure. Inset shows the absorption spectra of NK in presence of laponite deposited at 30 mN/m surface pressure.

Figure 6 Fluorescence spectra of mixed dyes (RhB18:NK) in laponite dispersion. Dye loading is 10% of CEC of laponite. Inset shows the enlarged spectrum of 100% RhB18 in laponite dispersion.

Figure 7a Fluorescence spectra of monolayer LB films of mixed dyes (RhB18:NK) in absence of laponite deposited at 15 mN/m surface pressure. Inset shows the enlarged spectra of LB films with mole fraction of RhB18 of 0.9 and 1.0.

Figure 7b Fluorescence spectra of monolayer LB films of mixed dyes (RhB18:NK) in presence of laponite deposited at 15 mN/m surface pressure. Inset shows the enlarged spectra of LB films with mole fraction of RhB18 of 0.5, 0.9 and 1.0. Dotted line shows the spectra of (NK + laponite) monolayer deposited at 30 $mNm^{-1}$ surface pressure.

Figure 7c Fluorescence spectra of monolayer LB films of mixed dyes (RhB18:NK) with laponite deposited at 15 mN/m surface pressure. Amount of NK is fixed at 20 $\mu l$ for all cases. The number denote corresponding RhB18 loading in $\mu l$. Inset shows the corresponding excitation spectra. The monitoring wavelength was 590 nm.

Table 1: Compressibilities of the monolayers of mixed dyes in the presence and absence of laponite.

Table 2: Absorption and fluorescence band maxima of NK
Table 3: Absorption and fluorescence band maxima of RhB18

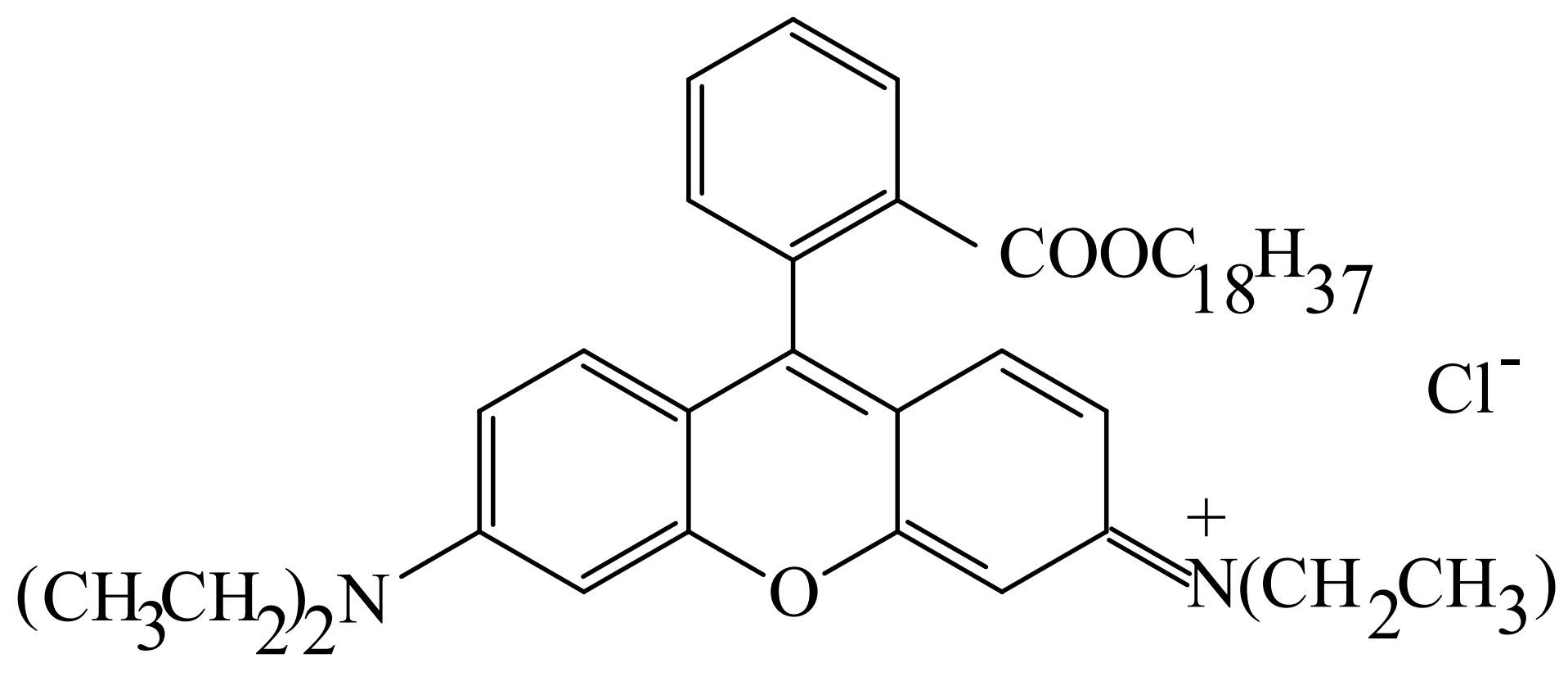

Figure 1(a) S. A. Hussain et. al.

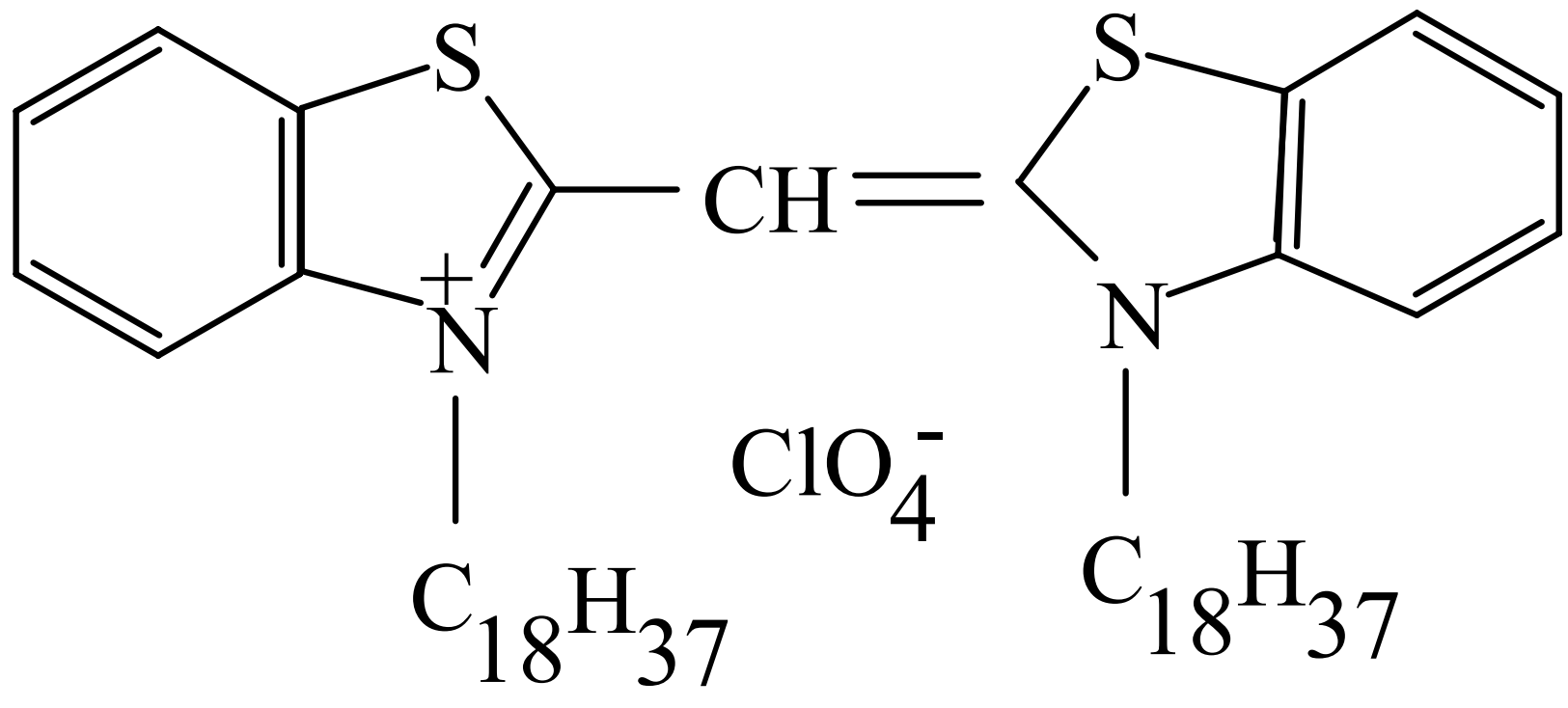

Figure 1(b) S. A. Hussain et. al.

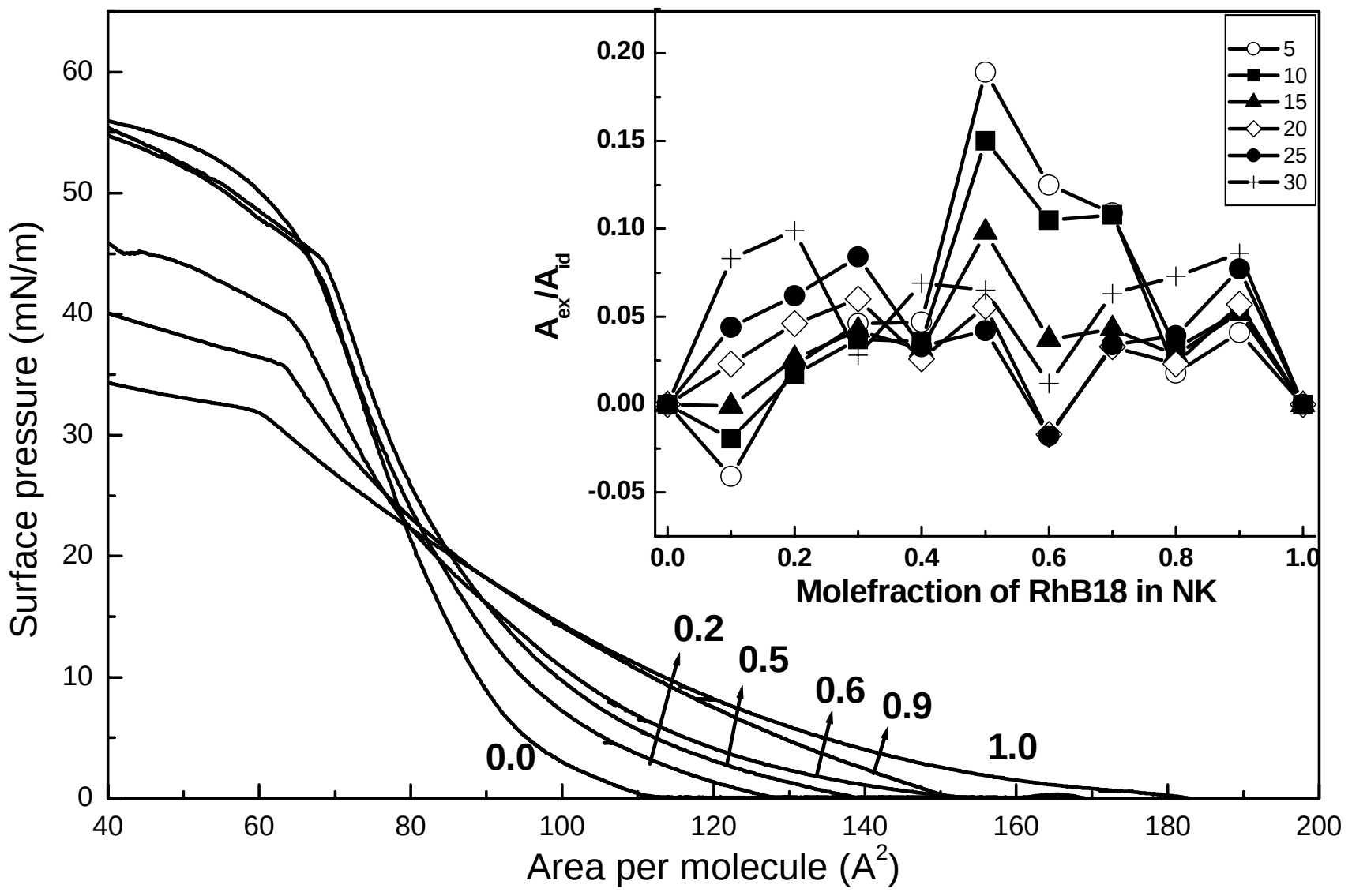

Figure 2a S. A. Hussain et. al.

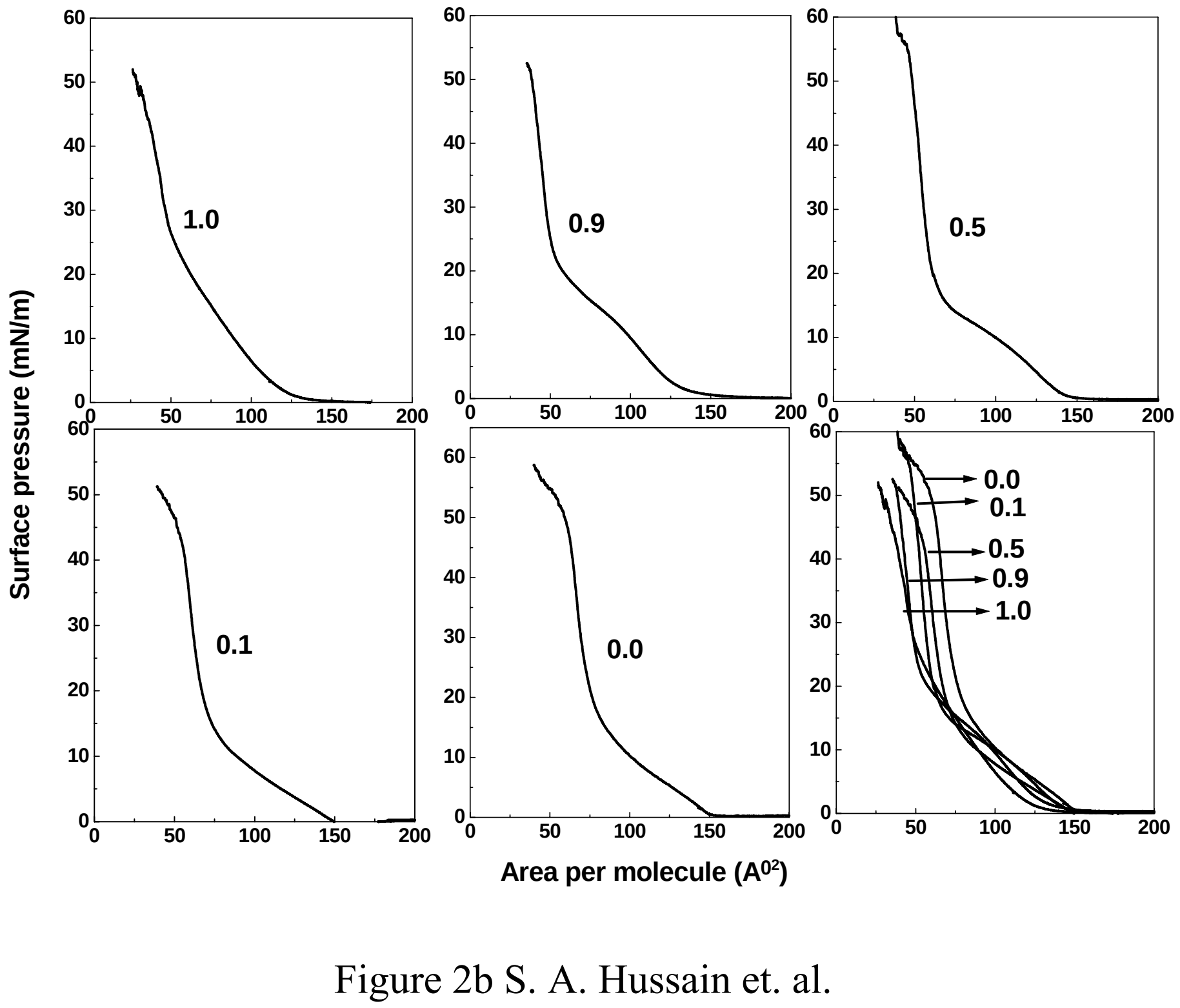

Figure 2b S. A. Hussain et. al.

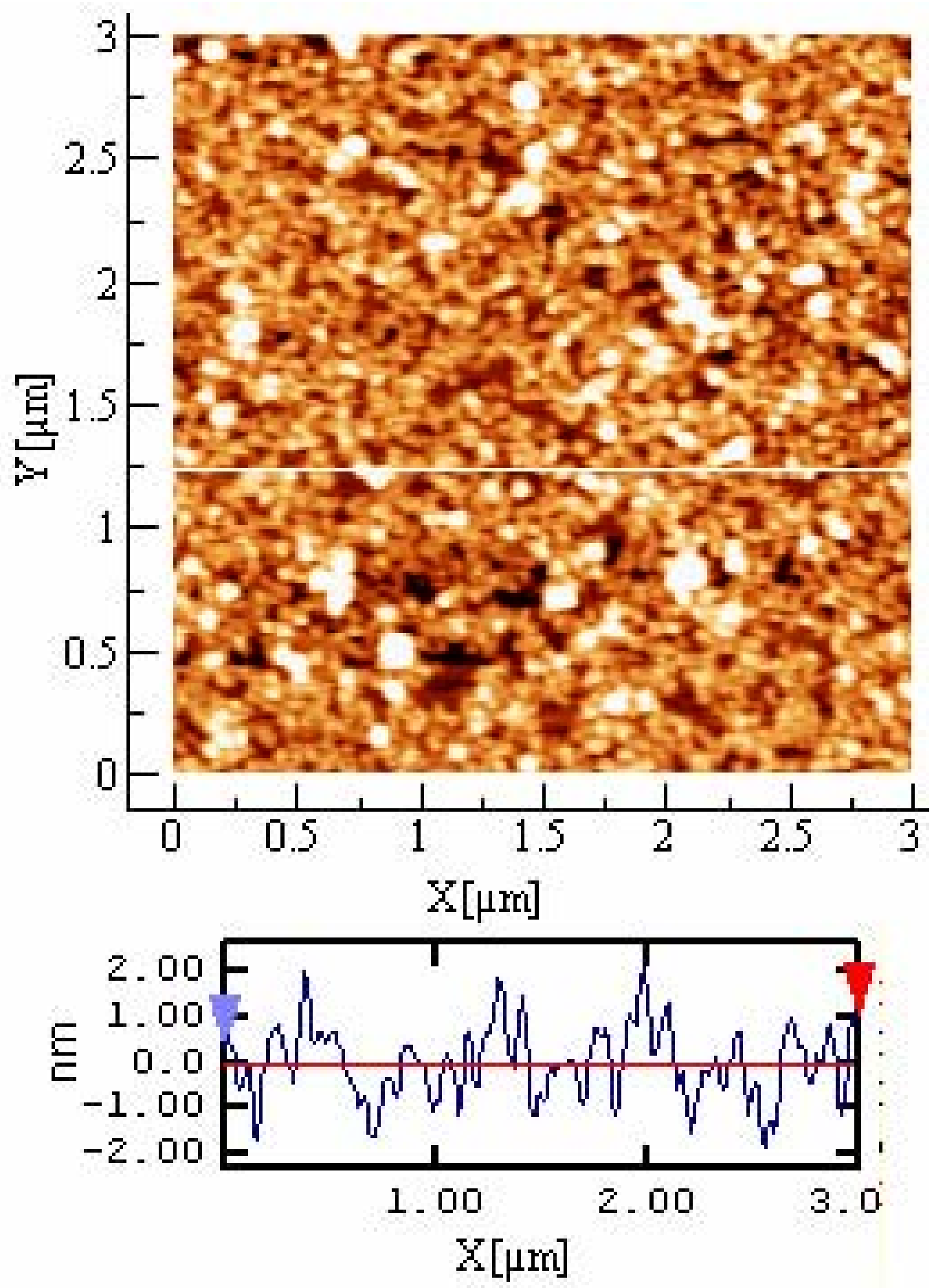

Figure 3 S. A. Hussain et. al.

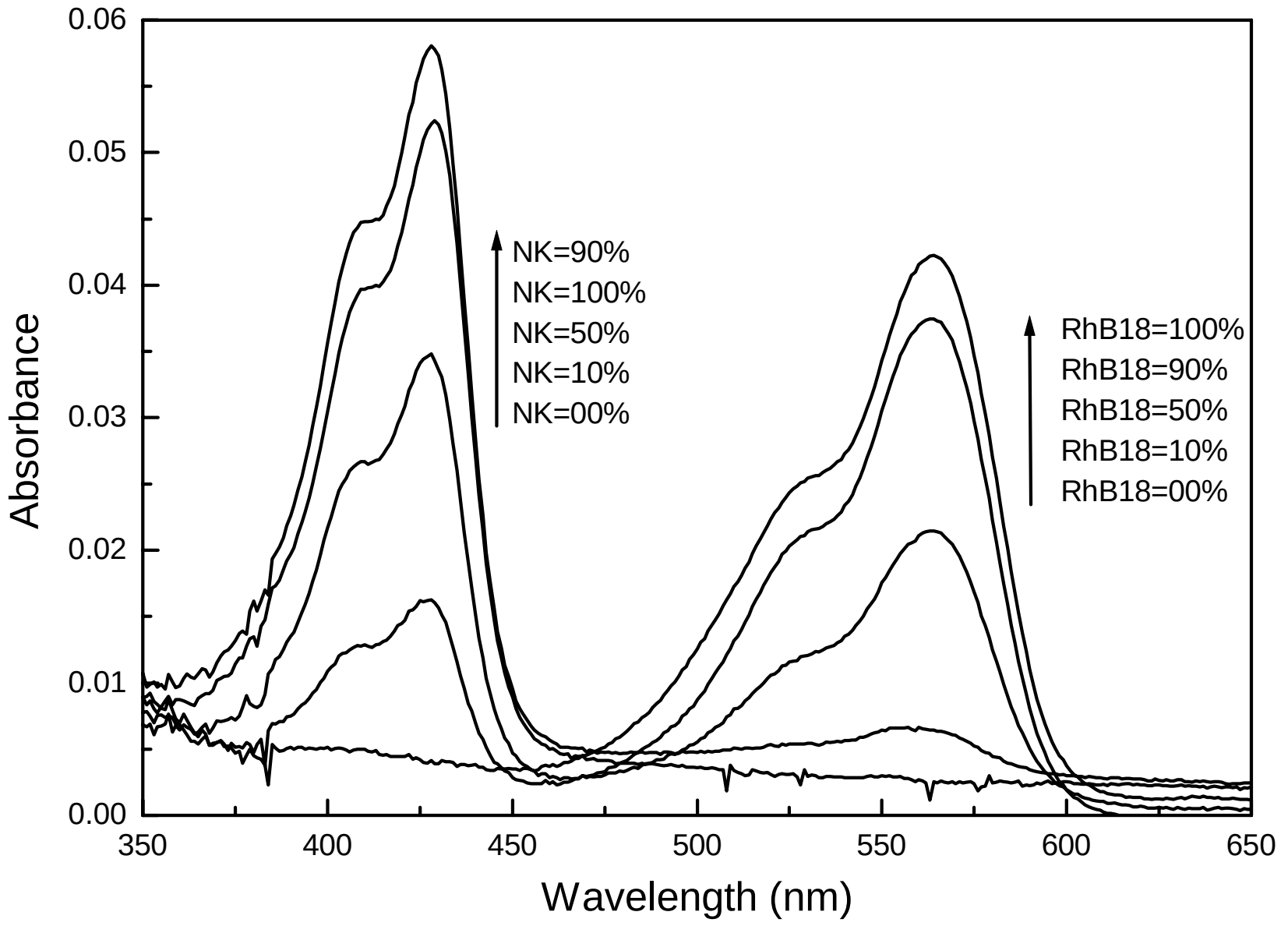

Figure 4 S. A. Hussain et. al.

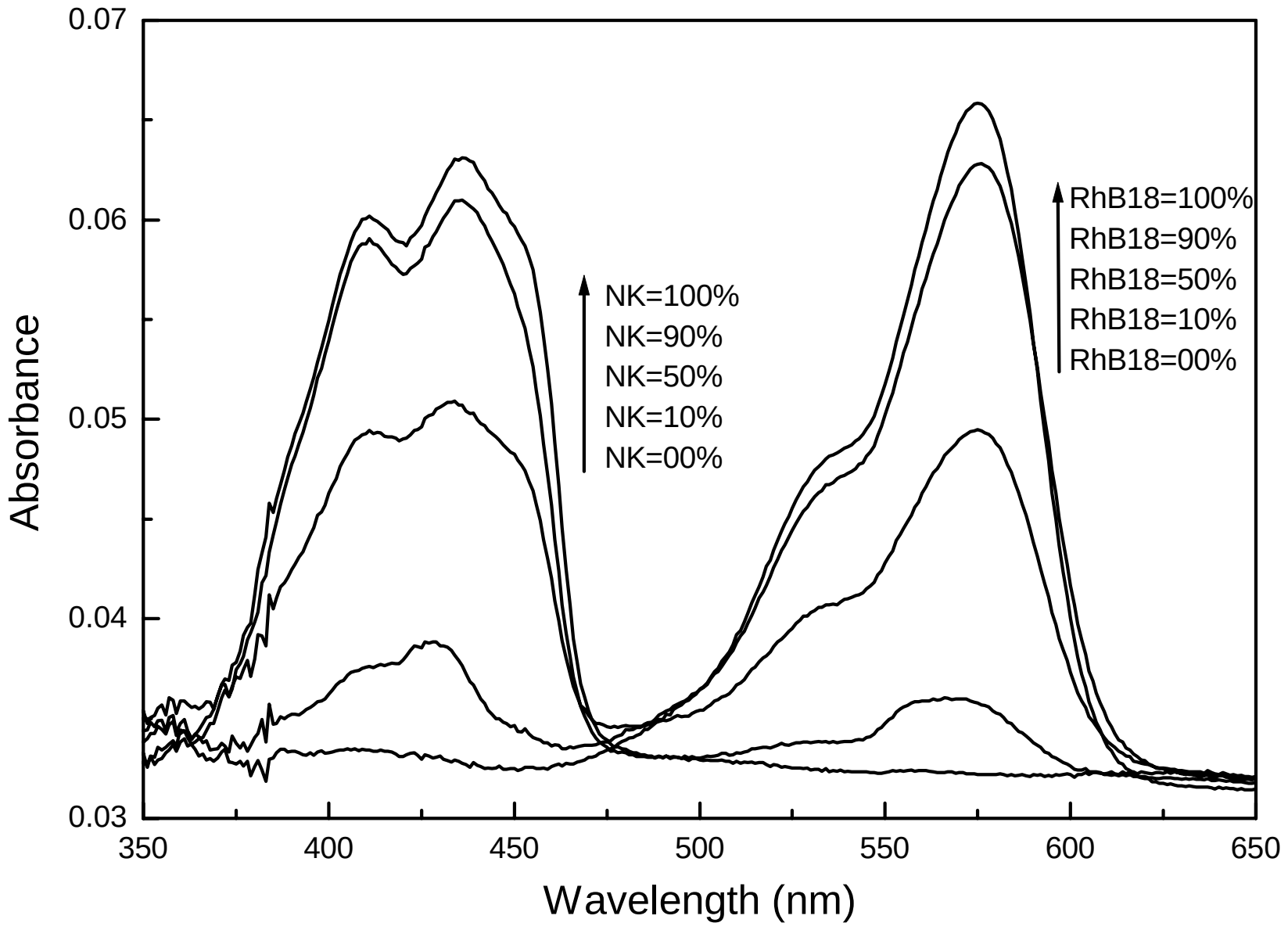

Figure-5a S. A. Hussain et. al.

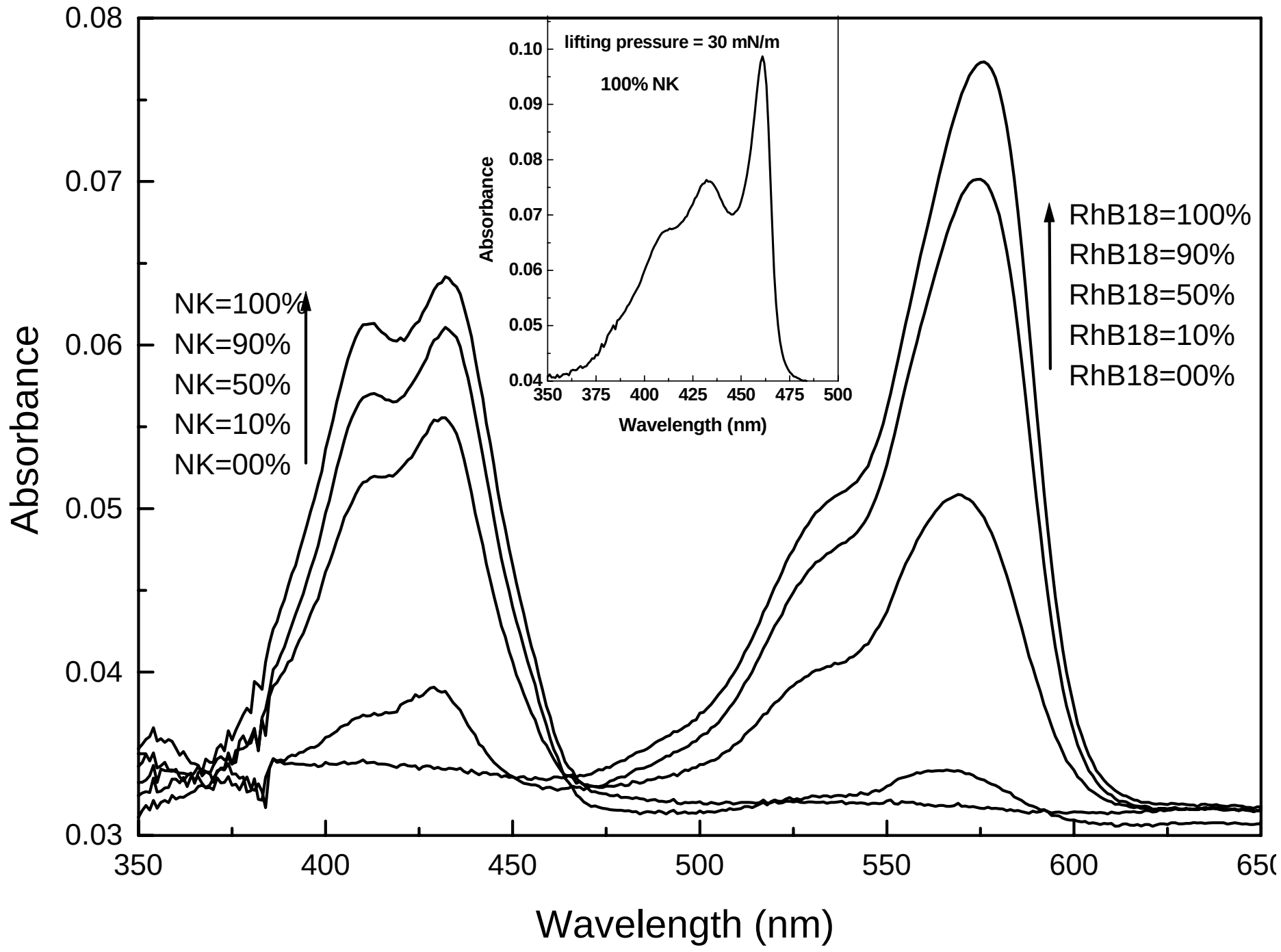

Figure 5b S. A. Hussain et. al.

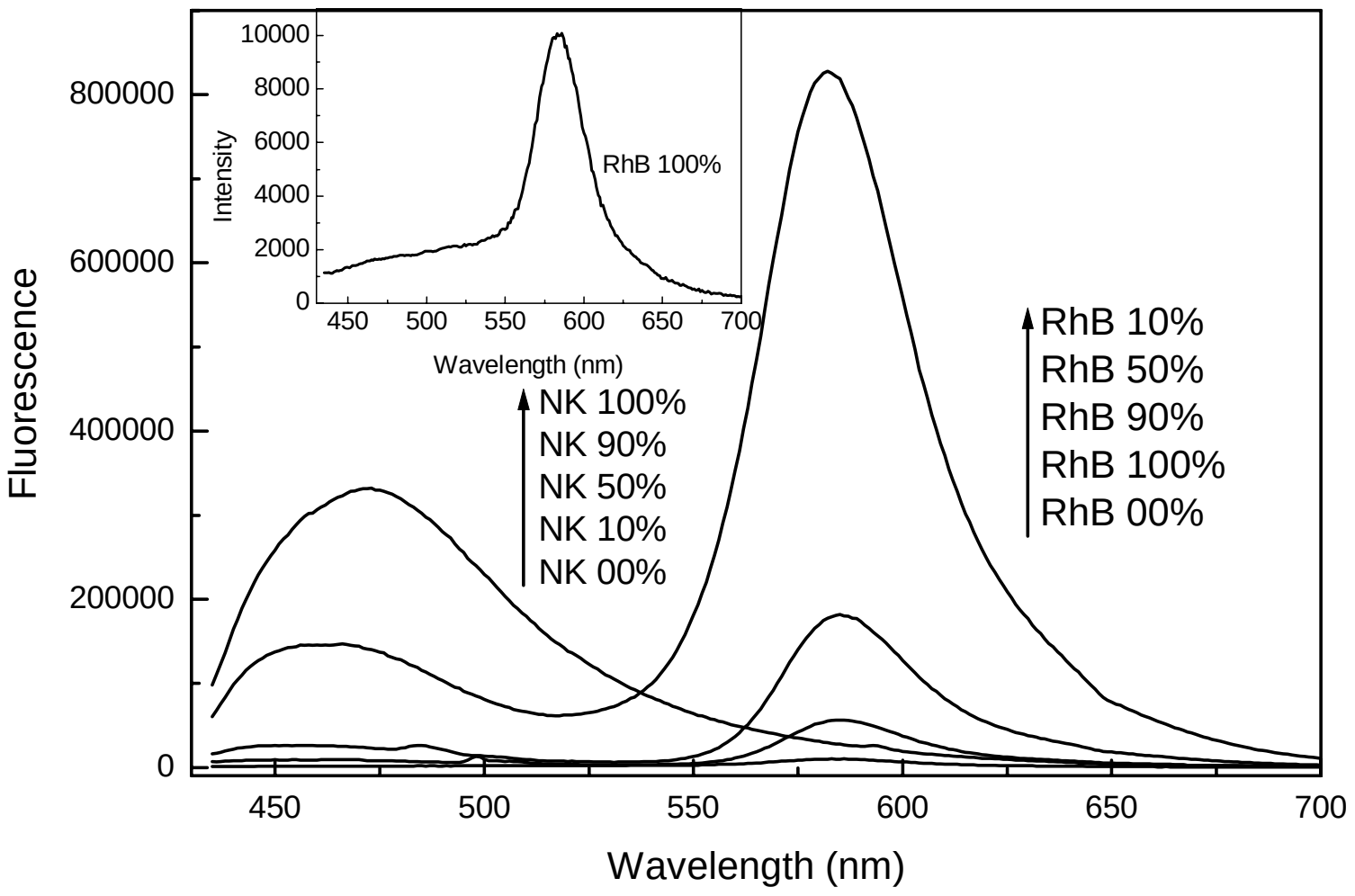

Figure 6 S. A. Hussain et. al.

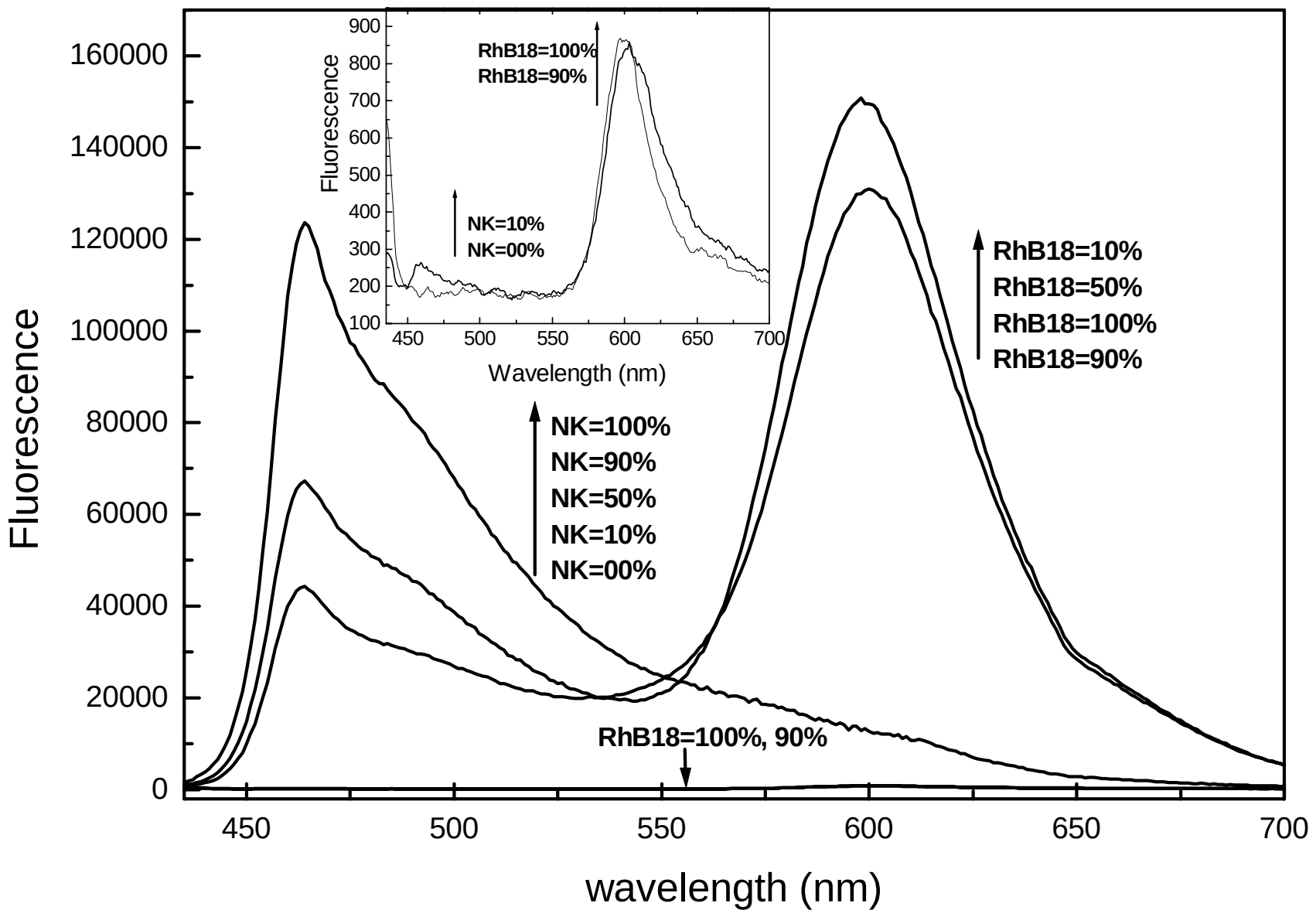

Figure 7a S. A. Hussain et. al.

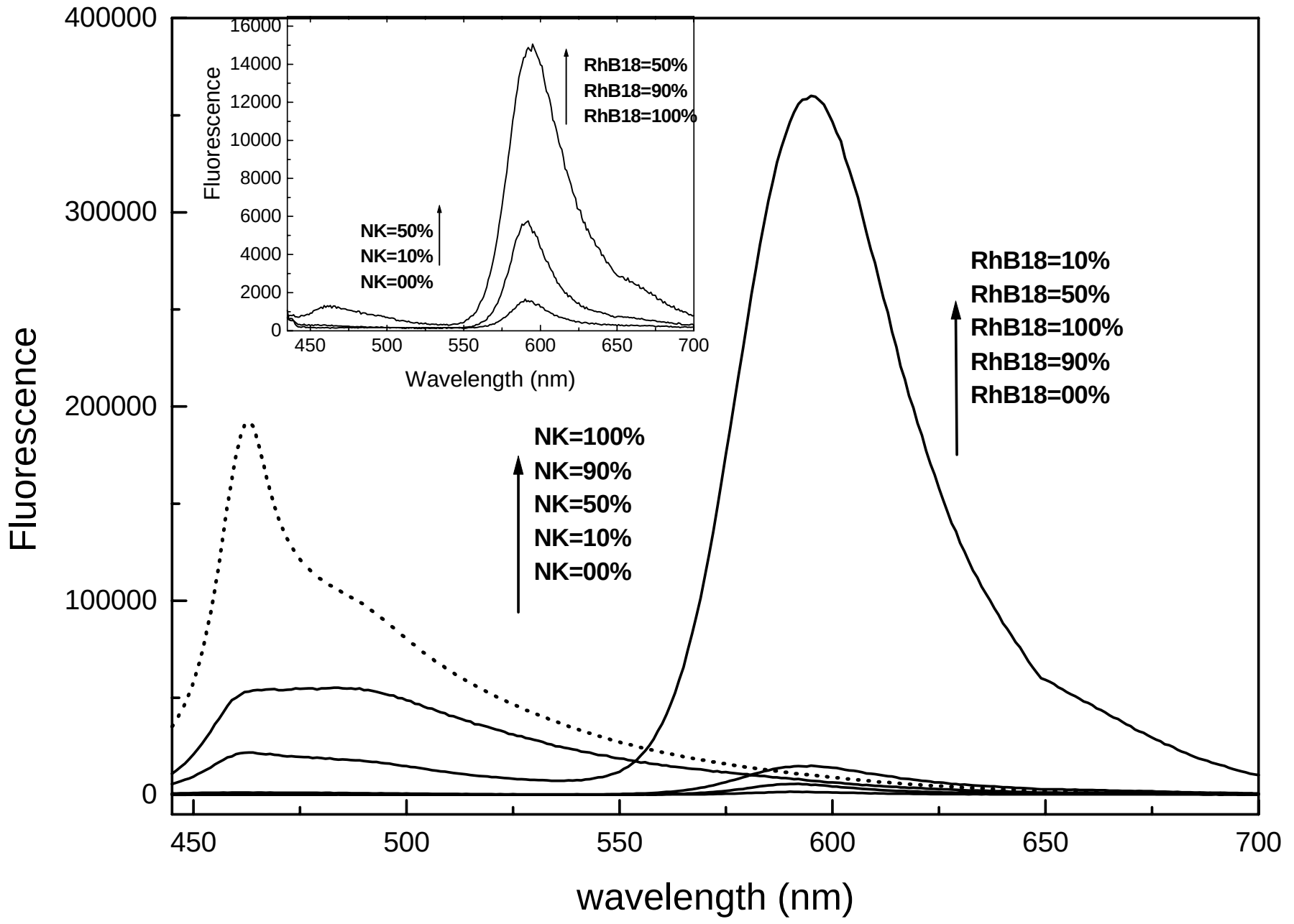

Figure 7b S. A. Hussain et. al.

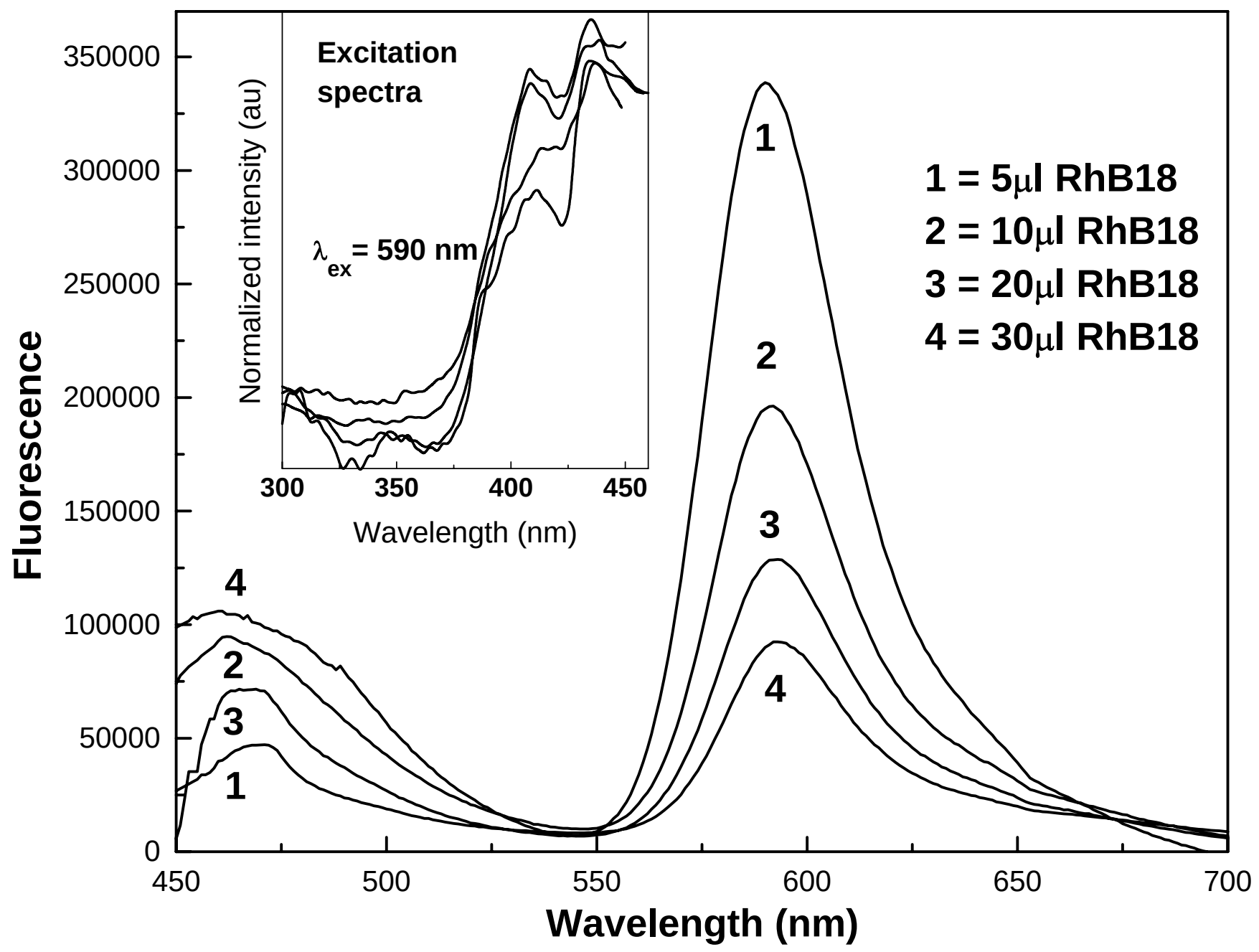

Figure 7c S. A. Hussain et. al.

Table 1: Compressibilities of the monolayers of NK and RhB18 at the air-water interface in the presence and absence of laponite sheets.

| Mole fraction of RhB18 | Compressibility ($mN^{-1}$) | | |
|---|---|---|---|
| | $C_1 = -\frac{1}{a_5}\frac{a_{10} - a_5}{10 - 5}$ | | $C_2 = -\frac{1}{a_{30}}\frac{a_{40} - a_{30}}{40 - 30}$ |
| | without clay | with clay | with clay |
| 0.0 | 13.5 | 40.2 | 5.6 |
| 0.1 | 16.4 | 47.6 | 7.2 |
| 0.2 | 20.5 | | |
| 0.3 | 23.0 | | |
| 0.4 | 24.9 | | |
| 0.5 | 30.3 | 38.1 | 5.9 |
| 0.6 | 29.4 | | |
| 0.7 | 27.8 | | |
| 0.8 | 26.6 | | |
| 0.9 | 28.1 | 29.4 | 8.6 |
| 1.0 | 31.2 | 30.6 | 15.4 |

Table 2: Spectroscopic signatures of NK

| | | monomer | | H- aggregate | | J- aggregate | |
|---|---|---|---|---|---|---|---|
| | | abs/nm | fluo/nm | abs/nm | fluo/nm | abs/nm | fluo/nm |
| Chloroform solution | | 430 | 483 | - | - | - | - |
| aqueous laponite dispersions[1] | | 427 | - | 406 | - | - | - |
| NK monolayer[1] | | 435 | - | 410 | - | 453 | 463 |
| NK laponite monolayer | 15 $mNm^{-1}$ | 435 | 490 | 410 | - | | 463 |
| | 30 $mNm^{-1}$ | 435 | 490 | 410 | - | 461 | 463 |

*[1]in these cases the fluorescence spectra are broad 450-490 nm, encompassing at least two components.*

Table 3: Spectroscopic signatures of RhB18

| | monomer | |
|---|---|---|
| | abs / nm | fluo / nm |
| Chloroform solution | 556 | 577 |
| aqueous laponite dispersions | 565 | 582 |
| NK monolayer | 576 | 599 |
| NK laponite monolayer | 565[1] – 576[2] | 595 |

*[1] at low loading: [2] at high loading*